\documentclass[12pt]{article}
\usepackage{amsmath, amssymb, epsfig, natbib}
\usepackage{amsmath, bbm, amssymb}
\usepackage{amsthm}
\usepackage{comment}
\usepackage{tabularx}
\usepackage{graphicx, graphics,color}
\usepackage{array}
\usepackage{caption}
\usepackage{booktabs}
\usepackage{float}
\usepackage{graphicx} 
\usepackage{subcaption} 
\usepackage{tabularx}   

\usepackage{pdflscape}
\usepackage{enumitem, xcolor}
\usepackage{multirow}
\usepackage{hyperref}
\usepackage{soul} 

\hypersetup{
	colorlinks=true,        
	citecolor=blue,         
	linkcolor=blue,         
	urlcolor=black,          
	bookmarksnumbered=true,  
	bookmarksopen=true,     
	breaklinks=true,        
	pdfstartview=FitH
}
\usepackage{epstopdf, mathrsfs, framed}
\usepackage{mathtools}
\usepackage{diagbox} 

\colorlet{linkequation}{blue}
\usepackage{algorithmic}
\usepackage[ruled,linesnumbered,vlined]{algorithm2e}

\newcommand{\bm}[1]{\mbox{\boldmath$ #1 $\unboldmath}}
\usepackage{algorithmic}
\usepackage[section]{placeins}
\usepackage{thmtools}
\usepackage{xpatch}

\def\prob {{\rm Pr}}
\def\E{\mathbb{E}}

\def\logit{\mbox{\rm logit}}

\makeatletter
\let\proof\thmt@original@proof
\xpatchcmd{\proof}{\itshape}{\bfseries}{}{}
\let\thmt@original@proof\proof
\makeatother

\date{}

\title{\vspace{-1cm} Consistency Assessment of Regional Treatment Effect for Multi-Regional Clinical Trials in the Presence of Covariate Shift}

\author{Kunhai Qing$^{1}$, Xinru Ren$^{1}$, Jin Xu$^{1,2, \ast}$, and Menggang Yu$^{3,\ast}$}

\begin{document}
	\maketitle
	\vspace{-.31cm}
	\noindent 
	{\small $^1$ School of Statistics, East China Normal University, Shanghai 200062, China\\
		$^2$ Key Laboratory of Advanced Theory and Application in Statistics and Data Science - MOE, East China Normal University, Shanghai 200062, China \\
		$^3$ Department of Biostatistics, University of Michigan School of Public Health, MI 48109, USA \\ 
		$^\ast$ Correspondence to: jxu@stat.ecnu.edu.cn; {menggang@umich.edu}
	}
	\vskip 1cm

\begin{abstract}  
Multi-Regional Clinical Trials (MRCTs) play a central role in the development of new therapies by enabling the simultaneous evaluation of drug efficacy and safety across diverse global populations. Assessing the consistency of treatment effects across regions is a fundamental aspect of MRCTs. Existing methods typically focus on region-specific marginal treatment effects. However, when treatment effect heterogeneity arises due to effect-modifying baseline covariates, distributional differences in these covariates can lead to erroneous conclusions. In this paper, we explicitly account for this phenomenon in the consistency assessment by considering the conditional average treatment effect. We propose a two-step assessment strategy that complements existing methods and mitigates the impact of treatment effect heterogeneity. Results from numerical studies demonstrate the effectiveness of the proposed approach.
\end{abstract}
    
\noindent {\bf Keywords: } Conditional Average Treatment Effect, Covariate Shift, Density 
Ratio, Multi-Regional Clinical Trial, Randomization
    
\section{Introduction}\label{sec:intro}    
With the increasing globalization of drug development, multi-regional clinical trials (MRCTs) are increasingly employed to enable faster accrual, more rapid patient access, improved collaboration and knowledge sharing, and greater regulatory harmonization. A major milestone in the development of MRCTs was the publication of the E17 guideline, titled ``General Principles for Planning and Design of Multi-Regional Clinical Trials'' by the International Council for Harmonisation of Technical Requirements for Pharmaceuticals for Human Use (ICH) in November 2017. The E17 guideline was subsequently adopted for regulatory implementation in Japan in June 2018, and in both the European Union and the United States in July 2018~\citep{Asano2021}.


While the overall outcomes of MRCTs are widely recognized as the primary basis for regulatory decisions, ensuring regional consistency remains critical. The United States usually serves as the primary market for pharmaceutical companies, leading to US participants often comprising the largest proportion in an MRCT, typically exceeding 60\%. Consequently, the proportions allocated to other regions are relatively small. Regulatory bodies, such as Japan's Pharmaceuticals and Medical Devices Agency and China's National Medical Products Administration, frequently require region-specific evidence to grant drug approvals for their respective markets. Indeed, E17 emphasizes sample size and statistical analysis planning to facilitate consistency evaluation.

However, as the worldwide drug development landscape is changing, this requirement for consistency assessment has become an important issue for the US regulatory agency as well. In the most recent FDA draft guidance on conducting MRCTs in oncology released on September 16, 2024 \citep{FDA2024}, the US FDA is concerned about the ``decreasing proportion of US participants included in oncology MRCTs" which ``can limit the assessment of treatment effect consistency between US enrolled participants and the effect observed for the overall study population in the MRCT". In particular, the guidance points out the phenomenon that ``the distribution of demographic characteristics or clinical characteristics of participants enrolled in these trials may differ significantly from the US population such that foreign data may not be appropriate to support an FDA regulatory decision".


This phenomenon is known as {\em covariate shift} in machine learning \citep{sugiyama2007covariate} and causal generalization \citep{ChenChenYu2023,ChenHuling2023biometrika}. A well-known MRCT example that drew significant attention to this issue is the PLATO (Platelet Inhibition and Patient Outcomes) trial \citep{NEJMoa0904327}. This Phase III study demonstrated that ticagrelor is superior to clopidogrel in the prevention of cardiovascular events in patients with acute coronary syndrome (ACS). However, for the North American region, the hazard ratio was in the wrong direction based on a pre-specified region-by-treatment interaction analysis \citep{CIRCULATIONAHA.111.047498}. 
Careful post hoc analysis revealed that a possible reason was the use of a higher dose of aspirin among trial subjects in the North American region. Justifiably, the nature of these two sets of analyses (i.e., pre-specified vs. post-hoc) resulted in a significant delay in the FDA's approval of ticagrelor \citep{CIRCULATIONAHA.121.055907}.     
Another MRCT study, which we will use for our numerical analysis, is the BELIEVE trial that supported the US FDA approval in November 2019 of luspatercept for the treatment of anemia in adult patients with $\beta$-thalassemia who require regular red blood cell transfusions \citep{CAPPELLINI2018163}. A stronger treatment effect was obtained in patients with lower baseline transfusion requirements \citep{bjh.18801}. As a result, luspatercept was approved in China only for patients who need regular transfusions of red blood cells and require $\le 15$ units per 24 weeks.

The E17 guideline points to descriptive summaries, graphical displays, covariate-adjusted analyses and tests of treatment-by-region interaction as possible strategies for such evaluation.  Although powerful for exploration and explanation, these strategies do not provide a pre-specified rule for consistency determination, especially in the presence of covariate shift.

Almost all existing literature on pre-specified procedures for consistency evaluation builds on two methods advocated in the white paper entitled ``Basic Principles on Global Clinical Trials" and released by Japan's Ministry of Health, Labor, and Welfare (MHLW) in September 2007 ~\citep{MHLW:2007}. In particular, the first method evaluates consistency by calculating the likelihood that the treatment effect within a specific region surpasses a fraction of the overall effect, assuming the overall effect is statistically significant. The second calculates the probability that treatment effects in all regions have the same direction. Both methods provide a pre-specified threshold for concluding consistency and hence can be used for sample size determination for regional allocation and as a planned analysis. They are indeed widely used in MRCTs \citep{Asano2021}.


Following the publication of the MHLW guidance, numerous subsequent methods \citep{ko_criterion, ikeda_sample_2010, tsou_consistency_2011, chen_decision_2012, tsong_assessment_2012, wu2020regional} have been developed to assess regional consistency and determine regional sample size requirements. These methods presume a consistent underlying treatment effect across regions with a shared variance. To introduce greater flexibility in modeling regional variations, an extended framework incorporating region-specific treatment effects was developed which enabled modifications and extensions of the criteria outlined by the MHLW guidance \citep{MHLW:2007} and inspired the creation of other methods for consistency evaluation \citep{chen_assessing_2010, quan_sample_2010}. Additionally, hypothesis-testing-based methods \citep{tanaka_qualitative_2012, teng_unified_2017, li2024regional} and random effects models \citep{quan_empirical_2013} have been proposed for assessing consistency.

Currently, no pre-specified procedures exist for consistency evaluation in MRCTs that explicitly address potential distributional shifts in treatment effect modifiers, which is also referred to as covariate shifts in this paper. In Section~\ref{sec:ATE-based consistency}, we first revisit MHLW-type assessment methods~\citep{MHLW:2007}, which directly use average treatment effect (ATE) estimates from a region of interest for consistency evaluation. We refer to these as one-step methods, in contrast to our proposed methods in Section~\ref{sec:CATE-based consistency}, which introduce an additional step to enable consistency assessment under distributional shift in treatment effect modifiers. Section~\ref{sec:simu} evaluates the performance of our proposed methods through simulation studies, demonstrating their operating characteristics. Section~\ref{sec:application} illustrates the practical utility of the proposed methods using a real-world example. Section~\ref{sec:disc} concludes with a discussion of the findings and directions for future work.

\section{One-step consistency assessment}
\label{sec:ATE-based consistency}

We first introduce our notation and describe the MHLW-type assessment methods~\citep{MHLW:2007}, which directly use ATE estimates from a region of interest for consistency assessment. 
Our proposed methods incorporate an additional step to enable further assessment of consistency under distribution shift in treatment effect modifiers.

Let $T \in \{-1,1\}$ denote the treatment indicator, where $T = 1$ corresponds to the treatment group and $T = -1$ to the control group. We assume that $T$ is randomized with $\prob(T = 1) = \pi_1$ and $\prob(T = -1) = \pi_{-1} = 1 - \pi_1$. Let $Y$ denote the endpoint of interest, and let ${\bm X} = (X_1, \cdots, X_p)^{\top}$ be a $p$-dimensional vector of baseline covariates measured prior to treatment assignment. The conditional average treatment effect (CATE) is defined as $\Delta({\bm X}) = \E(Y \mid T = 1, {\bm X}) - \E(Y \mid T = -1, {\bm X})$, while the ATE is the unconditional expectation, given by $\delta = \E\{\Delta({\bm X})\} = \E(Y \mid T = 1) - \E(Y \mid T = -1)$.


The MRCT is assumed to involve $K$ regions. Let $R$ represent the region. For each region $r=1,\cdots,K$, let $F_{r}(\cdot) = P({\bm X}\le \cdot \mid R=r)$ and $F_{-r}(\cdot) = P({\bm X}\le \cdot \mid R=-r)$ be the distribution functions of the covariate vector ${\bm X}$ in region $r$ and in regions outside $r$ ({or the complementary regions}) respectively, wherer the inequality here is interpreted component-wise. The region-specific CATEs are
$$\Delta_r({\bm X}) = \E(Y| T=1,{\bm X}, R=r) - \E(Y| T=-1,{\bm X}, R=r)$$
and 
$$\Delta_{-r}({\bm X}) = \E(Y| T=1,{\bm X}, R=-r) - \E(Y| T=-1,{\bm X}, R=-r).$$
Their corresponding ATEs are
$$\delta_r=\E_r\{\Delta_r({\bm X})\}\quad \text{and}\quad \delta_{-r}=\E_{-r}\{\Delta_{-r}({\bm X})\},$$
where the expectations are taken with respect to the regional distributions $F_{r}$ and $F_{-r}$.   
 
Let $\{(y_{i}, {\bm x}_{i}, t_{i}, r_{i}): i = 1,\cdots, n\}$ be a simple random sample of size $n$ of $(Y, {\bm X}, T, R)$, while $\mathcal{D}_{r}$ and $\mathcal{D}_{-r}$ denote the regional sample $\{(y_{i}, {\bm x}_{i}, t_{i}, r_{i}=r)\}$ and the { complementary} regional sample $\{(y_{i}, {\bm x}_{i}, t_{i}, r_{i}\neq r)\}$ respectively. For $r=1,\cdots,K$, let $n_{r}$ be the sample size of region $r$. Therefore the total sample size is $n=n_1+\cdots+n_K$. 
Let $\rho_{r}=n_{r}/n$ denote the regional proportion. For notational simplicity, we have assumed that
\begin{equation*}
        \frac{\sum_{i=1}^{n}I(t_{i}= 1)I(r_{i}= r)}{\sum_{i=1}^{n}I(t_{i}= 1)} = \frac{\sum_{i=1}^{n}I(t_{i}= -1)I(r_{i}= r)}{\sum_{i=1}^{n}I(t_{i}= -1)}=\rho_{r}. 
\end{equation*}
This condition is not necessary for our methodology development. 
The empirical estimator of the global ATE is
\begin{equation*}\label{eq:empirical-D}
        \widehat{\delta} = \frac{\sum_{i=1}^{n}y_{i} I(t_{i}=1)}{\sum_{i=1}^{n}I(t_{i}=1)} - \frac{\sum_{i=1}^{n}y_{i}I(t_{i}=-1)}{\sum_{i=1}^{n}I(t_{i}=-1)}.
\end{equation*}
Let $\widehat{Z}=\widehat{\delta}/\widehat{\rm se}(\widehat{\delta})$, where $\widehat{\rm se}(\widehat{\delta})$ denotes the standard error of $\widehat{\delta}$. We consider a superiority trial where a higher value of the response is preferred. The investigators shall reject the null hypothesis ($H_0: \delta=0$) of the overall treatment effect when $\widehat{Z}> z_{\alpha}$, where $\alpha$ is the nominal size and $z_{\alpha}$ is the upper $\alpha$ percentile of the standard normal distribution.

The empirical estimators of ATE for region $r$ and regions outside $r$ are respectively
\begin{align*}\label{eq:empirical-D-r}
\widehat{\delta}_{r}  & = \frac{\sum_{i=1}^{n}y_{i}I(t_{i}=1)I(r_{i}=r)}{\sum_{i=1}^{n}I(t_{i}=1)I(r_{i}=r)} - \frac{\sum_{i=1}^{n}y_{i}I(t_{i}=-1)I(r_{i}=r)}{\sum_{i=1}^{n}I(t_{i}=-1)I(r_{i}=r)}, \\
\widehat{\delta}_{-r}  & = \frac{\sum_{i=1}^{n}y_{i}I(t_{i}=1)I(r_{i}\neq r)}{\sum_{i=1}^{n}I(t_{i}=1)I(r_{i}\neq r)} - \frac{\sum_{i=1}^{n}y_{i}I(t_{i}=-1)I(r_{i}\neq r)}{\sum_{i=1}^{n}I(t_{i}=-1)I(r_{i}\neq r)}.
\end{align*}

Traditional consistency criteria consider a regional treatment effect to be consistent with the overall treatment effect when the regional ATE, $\delta_r$, is equal to (fixed effects model) or approximately equal to (random effects model, under a common prior distribution) the overall ATE, $\delta$, or the ATE outside region $r$, $\delta_{-r}$. In contrast to the formal hypothesis testing framework, the MHLW-type assessment methods~\citep{MHLW:2007} focus on ensuring a high probability that the event $\{\widehat{\delta}_{r} > q \widehat{\delta} \}$ \citep{quan_sample_2010} or  $\{ \widehat{\delta}_{r} > q \widehat{\delta}_{-r} \}$ \citep{ko_criterion} occurs, where $q\ge 0.5$ is
the fraction of the overall treatment effect to preserve. This is also known as `selection design' \citep{Gibbons:Olkin:Sobel:99, YiYu2024, Sargent:Goldberg:01}. These criteria rely solely on marginal treatment effects and do not account for covariates; therefore, we classify them as marginal consistency criteria.

Regulatory agencies primarily require that an MRCT demonstrate overall success. Consistency is considered only after overall statistical significance has been achieved. Therefore, it is both reasonable and appropriate to assess consistency as a \textit{conditional event}. In practice, many consistency assessment methods follow this principle, evaluating consistency only when the overall trial reaches statistical significance; see \citet{Kawai:2008}, \citet{ko_criterion}, \citet{wu2020regional} and references therein. Thus we may rewrite the above events as 
\begin{equation}\label{eq:conditional-events}
    \left\{\dfrac{\widehat{\delta}_{r}}{\widehat{\delta}} > q \, \Big \vert \,  \widehat{Z}> z_{\alpha}\right\}\text{ and } \left\{ \dfrac{\widehat{\delta}_{r}}{\widehat{\delta}_{-r}} > q  \, \Big \vert\, \widehat{Z}> z_{\alpha}\right\},
\end{equation} with the latter equivalent to
\begin{eqnarray*} 
        \left\{  \dfrac{\widehat{\delta}_{r}}{\widehat{\delta}} > \dfrac{q}{1-\rho_{r}+q \rho_{r}} \, \Big \vert\, \widehat{Z}> z_{\alpha}\right\}.
\end{eqnarray*} 
Once this conditional event occurs, consistency is claimed. Hereafter, we use the second event to explain our method.

\section{Two-step consistency assessment}
\label{sec:CATE-based consistency}
For any specific region $r$, the regional ATE is defined as $\delta_r = \E_r\{\Delta_r({\bm X})\} = \int \Delta_r({\bm x}) dF_r({\bm x})$. This expression highlights that differences in ATE between region $r$ and the overall population (or the region outside $r$) can arise from: (i) a substantial difference in CATE, i.e., $\Delta_r({\bm X}) \neq \Delta_{-r}({\bm X})$; (ii) a substantial difference in covariate distributions, i.e., $F_r({\bm X}) \neq F_{-r}({\bm X})$; or (iii) a combination of both factors.

As a result, when inconsistency is suspected based on the one-step assessment, i.e., when ${\widehat{\delta}_r}/{\widehat{\delta}} \le q$, it is important to further investigate the underlying cause. Specifically, one should assess whether the inconsistency arises from differences in CATE, covariate distribution, or both. To this end, we propose the following generic algorithm:
\begin{itemize}
        \item [(i)] Conduct the one-step consistency assessment. If the conclusion is consistency, stop;
        
        \item [(ii)] Otherwise, if the conclusion is inconsistency, evaluate whether $\Delta_{r}({\bm X})$ and $ \Delta_{-r}({\bm X})$ are close to each other;
        
\begin{itemize}
        \item [(ii.1)] If $\Delta_{r}({\bm X})$ and $ \Delta_{-r}({\bm X})$ are not close to each other, the final conclusion is inconsistency; stop;  

        \item [(ii.2)] If $\Delta_{r}({\bm X})$ and $ \Delta_{-r}({\bm X})$ are close to each other, estimate a covariate shift-adjusted ATE for region $r$ and re-evaluate consistency;  

        \item [(ii.3)] The conclusion based on (ii.2) is then the final conclusion.
\end{itemize}
\end{itemize}

This algorithm augments the one-step method with a salvage strategy based on Step (ii). The rationale behind Step (ii)---evaluating whether $\Delta_r({\bm X})$ and $\Delta_{-r}({\bm X})$ are similar---is as follows. It is conceivable that differences in both CATE and covariate distributions may counteract each other. That is, it is possible for $\delta_r = \delta_{-r}$ to hold even when $\Delta_r({\bm X}) \ne \Delta_{-r}({\bm X})$ and $F_r({\bm X}) \ne F_{-r}({\bm X})$ both hold. As a result, before estimating the covariate shift-adjusted ATE for region $r$, it is necessary to assess the similarity between $\Delta_r({\bm X})$ and $\Delta_{-r}({\bm X})$. Otherwise, it is not appropriate to proceed with the salvage strategy. We describe the steps as follows.

\subsection{Evaluating similarity between $\Delta_{r}({\bm X})$ and $\Delta_{-r}({\bm X})$ }\label{subsec:intrinsic-inconsistent}


Our strategy for evaluating whether $\Delta_{r}({\bm X})$ and $\Delta_{-r}({\bm X})$ are close to each other is through an estimate of individual treatment effect (ITE). In particular, it is easy to see that \citep{Liang03042022,ChenTianCaiYu}
$$
\E\biggl[\dfrac{T_i}{\pi_{T_i}} Y_{i} \,\bigg |\,  {\bm X}_{i}, R_i=r \biggr] = \Delta_r({\bm X}_i),$$ 
where $\pi_{T_i}=\pi_1$ when $T_i=1$ and $\pi_{T_i}=\pi_{-1}$ when $T_i=-1$.
Further, for any function $m({\bm X}_i, R_i)$, 
\begin{eqnarray}
\E\biggl [\dfrac{T_i}{\pi_{T_i}} \big\{Y_{i}-m({\bm X}_i, R_i)\big\} \,\bigg |\,  {\bm X}_{i}, R_i=r \biggr] = \Delta_r({\bm X}_i) \label{ITE}
\end{eqnarray} 
Therefore ${\pi_{T_i}}^{-1}{T_i} \big\{Y_{i}-m({\bm X}_i, R_i)\big\}$ can be considered as an estimated ITE. 
Consequently, we consider the following working model 
\begin{eqnarray} 
\dfrac{T_i}{\pi_{T_i}} \big\{Y_i-m({\bm X}_i, R_i)\big\} = \beta_{0} + {\bm\beta}_{X}^\top {\bm X}_i + \beta_{R}\cdot I(R_i=r) + {\bm\beta}_{R X}^\top {\bm X}_i \cdot I(R_i=r) + \varepsilon_i. \label{eq:working-model}
\end{eqnarray}
 
\citet{Wu:2018} showed that the most efficient choice for $m({\bm X}_i, R_i)$ is $\pi_{-1}\E[Y_i\mid T_i=1, {\bm X}_i, R_i] + \pi_1 \E[Y_i\mid T_i=-1, {\bm X}_i, R_i]$. Because the expectations are unknown, they need to be estimated. \citet{Wu:2018} proposed a Leave-One-Out Potential Outcomes (LOOP) estimator. By leveraging information from the other participants' covariates (${\bm x}_{j}$, $j\neq i$) to provide an estimate $\widehat{m}_{i}$, the LOOP estimator achieves a substantial reduction in variance. Furthermore, the LOOP estimator is robust as it does not rely on any model assumptions, which is particularly beneficial given the typically smaller sample size from region $r$.
Because a biased $\widehat{m}_{i}$ can be used in \eqref{eq:working-model} and the convergence rate of $\widehat{m}_{i}$ is not required to be $\sqrt{n}$, practitioners could use simple models such as linear regression, or sophisticated machine learning methods such as random forest \citep{Breiman:2001} to estimate it.
  
Finally, as ${\bm\beta}_{RX}$ is a vector, we use the Wald test or the likelihood ratio test for its significance. Rejection of $H_0: {\bm\beta}_{RX} = {\bm 0}$ indicates that $\Delta_r({\bm X}) \ne \Delta_{-r}({\bm X})$, that is, there exists a significant difference in the interaction of the covariates ${\bm X}$ on the endpoint $Y$ between region $r$ and regions outside $r$. Otherwise, we consider the two CATE functions to be similar.

\subsection{Covariate-shift adjusted ATE}\label{subsec:consistency-covariate-shifts}
When $\Delta_{r}({\bm X})$ and $\Delta_{-r}({\bm X})$ are very similar, a large difference between the ATEs $\delta_{r}$ and $\delta_{-r}$ should mainly come from distributional differences in treatment effect modifiers. Under the presumption that the consistency of the CATE should lead to a conclusion of the consistency of the ATE, we consider the following quantity, 
\begin{eqnarray*}
        \delta_{r}^{*} &=& \E_{r}\left\{ \Delta_{r}({\bm X})\cdot \frac{f_{-r}({\bm X})}{f_{r}({\bm X})} \right\} = \E_{-r}\left\{ \Delta_{r}({\bm X})  \right\}.
\end{eqnarray*} 
Therefore $\delta_{r}^{*}$ is the CATE of region $r$ marginalized according to the covariate distribution of region $-r$. We justify below that $\delta_{r}^{*}$ can be directly compared with $\delta_{-r}$, the ATE of the complementary regions, to evaluate the consistency of the treatment effect,
\begin{align*}
        \delta_{r}^{*} &= \E_{r}\left\{ \Delta_{r}({\bm X})\cdot \frac{f_{-r}({\bm X})}{f_{r}({\bm X})} \right\} \\ 
        &= \int \Delta_{r}({\bm x})\cdot \frac{f_{-r}({\bm x})}{f_{r}({\bm x})} f_{r}({\bm x}) {\rm d}{\bm x} \\ 
        &= \int \Delta_{-r}({\bm x})\cdot  f_{-r}({\bm x}) {\rm d}{\bm x} \\ 
        & = \E_{-r}\left\{ \Delta_{-r}({\bm X})  \right\}  =  \delta_{-r}
\end{align*}

The estimation of $\delta_{r}^{*}$ can be based on
\begin{eqnarray}
        \widehat{\delta}_{r}^{*}  &=& n_{r}^{-1}\sum_{i\in {\cal D}_r} \widehat{\Delta}_{r}({\bm x}_i)\cdot \frac{\widehat{f}_{-r}({\bm x}_i)}{\widehat{f}_{r}({\bm x}_i)} 
        \label{density ratio estimate}
\end{eqnarray} 
where $\widehat{\Delta}_{r}({\bm x}_i)$ is an estimate of the ITE. In particular, the LOOP estimator introduced in Section~\ref{subsec:intrinsic-inconsistent} can be used. The density ratio estimate ${\widehat{f}_{-r}({\bm x}_i)}/{\widehat{f}_{r}({\bm x}_i)}$ can be obtained using kernel smoothing methods \citep{Wandbook}.

When the dimension $p$ is moderately large, the density ratio estimate ${\widehat{f}_{-r}({\bm x}_i)}/{\widehat{f}_{r}({\bm x}_i)}$ can be unstable. 
One strategy to deal with this is to use sufficient dimension reduction \citep{Liang03042022}. In particular, when ${\Delta}_{r}({\bm x}_i)$ depends on only a few linear combinations ${\bm x}^d_i\equiv {\bm \alpha}^\top {\bm x}_i$ where ${\bm \alpha}$ is a matrix of dimension $p\times d$ for some small $d$, then the density ratio of ${\bm x}^d_i$ between the two regions can be used in \eqref{density ratio estimate}. One can even fix $d=1$ and use a single index model \citep{Xiong03072017} to estimate ${\Delta}_{r}({\bm x}_i)$.

However, linear combinations may not be easy to interpret. Therefore, we consider the following alternative. For each component $x_s$ of ${\bm x}$, $s=1,\cdots,p$, we define a new estimate as:
\[  
\widehat{\delta}_{r,s}^{*}= n_{r}^{-1} {\sum_{i\in\mathcal{D}_{r}}\widehat{\Delta}_{r}({\bm x}_i)\cdot \dfrac{\widehat{f}_{-r,s}({x}_{s,i})}{\widehat{f}_{r,s}({x}_{s,i})}}
\]
We can see that $\widehat{\delta}^{*}_{{r,s}}$ converges to
\begin{eqnarray}
&& \mathbb{E}_r \bigg\{ \Delta_{r}({\bm X})\cdot \dfrac{{f}_{-r,s}(X_s)}{{f}_{r,s}(X_s)} \bigg\} \nonumber \\
&&= \int \Delta_{r}({\bm x})\cdot \dfrac{{f}_{-r,s}(x_s)}{{f}_{r,s}(x_s)} \cdot {f}_{r}({\bm x}) {\rm d} {\bm x} \nonumber  \\
&&= \int \Delta_{r}({\bm x})\cdot {f}_{r}({\bm x}_{-s}\mid x_s) \cdot {{f}_{-r,s}(x_s)}{\rm d} {\bm x} \nonumber \\
&&=\int \bigg\{ \int \Delta_{r}({\bm x})\cdot {f}_{r}({\bm x}_{-s}\mid x_s) {\rm d} {\bm x}_{-s} \bigg\} \cdot {{f}_{-r,s}(x_s)} {\rm d}x_s \label{eq:conditional-s}
\end{eqnarray}
where ${\bm x}_{-s}$ are the components of ${\bm x}$ excluding $x_s$. The quantity in the braces above is
\begin{eqnarray*}
        \mathbb{E}_r\big\{ \Delta_r({\bm X})\mid X_s=x_s \big\} = \Delta_r(x_s) , 
\end{eqnarray*} where
$$\Delta_r(x_s) \equiv \E(Y| T=1,X_s=x_s, R=r) - \E(Y| T=-1,X_s=x_s, R=r)$$ is the treatment effect conditioning only on $X_s$. 
                
Consequently, \eqref{eq:conditional-s} equals 
$$
\int \Delta_r(x_s) {{f}_{-r,s}(x_s)} {\rm d}x_s   = \mathbb{E}_{-r} \big\{ \Delta_r(X_s) \big\} \equiv \delta^{*}_{{r,s}}.
$$

Under the assumption of similar CATEs across regions, $\Delta_r(x_s)$ and $\Delta_{-r}(x_s)$ should also be similar if we further assume $f_{r}(\bm x_{-s}|x_s)$=$f_{-r}(\bm x_{-s}|x_s)$. Consequently, $ \mathbb{E}_{-r}\{\Delta_r(X_s)\}$ will be similar to  $\mathbb{E}_{-r}\{\Delta_{-r}(X_s) \}= \delta_{-r}$. This property makes $\delta^{*}_{{r,s}}$ a suitable target for our assessment.
Having calculated $\widehat{\delta}_{r,s}^{*}$, for $s = 1, \cdots, p$, we propose a modified version of the second conditional event in \eqref{eq:conditional-events}, given by: 
\begin{equation} 
     \left\{ \max_s \bigg(\dfrac{ \widehat{\delta}_{r,s}^{*}}{\widehat{\delta}_{-r}} \bigg) > q^{(2)} \, \Big \vert \, \widehat{Z}> z_{\alpha}\right\}, \label{eq:CATE-adjusted-event}
\end{equation} 
where $q^{(2)}$ is the threshold in the second step.


Let $q^{(1)}$ denote the fraction of overall treatment effect to preserve the first step of our proposed two-step method. By design, when $q^{(1)}=q$ (i.e., the same threshold as the one-step method), our propose two-step method yields no smaller CP than the one-step method regardless of the value of $q^{(2)}$. In light of the approach of hypothesis testing, we seek a combination of $q^{(1)}$ and $q^{(2)}$ such that, when there is no covariate shift, it attains lower CP (analogous to type I error) than the one-step method given that the CATE functions are different, and achieves higher CP (analogous to power) given that the CATE functions are similar. Therefore, we intentionally choose a larger $q^{(1)}$ than $q$ to guarantee the consistency claimed by the first step with high confidence, and use $q^{(2)}=q$ to rectify the false claim of inconsistency due to actual covariate shift. We explore different choices of $q^{(1)}$ through simulation in Section~\ref{sec:simu}.

\subsection{Algorithm for two-step assessment and practical consideration}\label{subsec:combined-consistency-assessment}
The full procedure for the two-step assessment in determining consistency between region $r$ and the other regions is detailed in Algorithm~\ref{algorithm:combined-consistency-assessment}. 

\begin{algorithm}[t]
        \caption{Two-step consistency assessment}\label{algorithm:combined-consistency-assessment}
        \KwIn{$\alpha,q^{(1)},q^{(2)},\mathcal{D}_{r}$ and $\mathcal{D}_{-r}$  }
        \KwOut{``Consistency" or ``Inconsistency"}
        Calculate $\widehat{Z}$, $\widehat{\delta}_{r}$, $\widehat{\delta}_{-r}$ and $\widehat{\delta}^{*}_{r,s},s=1,\cdots,p$\;

        \If{$\widehat{Z}> z_{\alpha}$ and $\widehat{\delta}_{r}/\widehat{\delta}_{-r} > q^{(1)}$ (i.e., the marginal consistency assessment)}{
        \Return{``Consistency"}\;
    }
    \Else{
                \If{$H_0: \Delta_r({\bm X}) = \Delta_{-r}({\bm X})$ is rejected}{
        \Return{``Inconsistency"}\;
        }
                \Else{
                                \If{At least one of the inequalities $\widehat{\delta}^{*}_{r,s}/\widehat{\delta}_{-r} > q^{(2)}, s = 1, \cdots, p$, holds}{
                                \Return{``Consistency"}\;
                                \Else{\Return{``Inconsistency"}\;}
                        } 
        } 
    }
\end{algorithm}

The augmentation step used to claim consistency in the algorithm may raise the concern that it could increase false positive claims when the scenarios described at the beginning of Section~\ref{sec:CATE-based consistency} do not hold, specifically, when there is no covariate shift or when the CATE does not depend on a particular covariate $X_s$. Here we argue that, from a theoretical standpoint, such a concern is not warranted.

First, if there is no distributional shift for a particular covariate $X_{s}$, the density ratio ${f_{-r,s}}/{f_{r,s}}$ equals one. As a result, the estimand $\delta^{*}_{r,s}$ reduces to $\delta_{r}$. Similarly, when $\Delta_r(x_s)$ does not depend on $x_s$, then it equals $\delta_r$ which also implies that $\delta^{*}_{r,s}=\delta_{r}$. In both cases, the ratio $\widehat{\delta}^{*}_{r,s}/\widehat{\delta}_{-r}$ used in the augmentation step will be very close to the ratio $\widehat{\delta}_{r}/\widehat{\delta}_{-r} $ from the first step. Therefore, the operating characteristics of the traditional one-step method are preserved. This behavior is also supported by our simulation studies.

On the other hand, if the inequality $\widehat{\delta}^{*}_{r,s}/\widehat{\delta}_{-r} > q^{(2)}$ holds for some $s = 1, \cdots, p$, we can conclude that the observed discrepancy between the regional ATE and the complementary regional ATE is attributable to a distribution shift in that specific covariate. This enables the identification of the contributing covariate (i.e., effect modifier) and enhances the interpretability of the findings. Therefore, the set of ratios
$\{\widehat{\delta}^{*}_{r,s}/\widehat{\delta}_{-r}, s = 1, \cdots, p\}$ provides useful information about the influence of individual covariates on ATE discrepancies. Ranking these ratios can be informative and may serve as an additional visualization tool for overall consistency assessment, in line with the E17 guideline.   

In MRCTs, interpreting the CATE can be challenging, particularly when covariates are continuous. A common and effective approach is to discretize continuous covariates into categorical variables. This transformation renders the unknown function $\Delta({\bm X})$ piecewise constant, which significantly improves both the interpretability and the effectiveness of subsequent statistical analyses, such as ANCOVA. Accordingly, when applying our proposed consistency assessment method, we recommend discretizing the covariates, as this also simplifies the estimation of the density ratio.

While our proposed method is readily applicable to continuous and binary outcomes, significant methodological challenges arise when dealing with time-to-event endpoints. The CATE framework is not directly applicable to survival data. Moreover, a common alternative, the conditional hazard ratio, poses difficulties for consistency assessment because its expectation generally does not equal the overall hazard ratio.

To circumvent this limitation, we utilize the \textit{Restricted Mean Survival Time} (RMST) \citep{Pepe:1989} as a more suitable primary metric for treatment effect in this context. RMST measures the survival time up to a clinically relevant truncation point $\tau$ and serves as a compelling alternative to the hazard ratio \citep{Uno:2014}. To address the issue of right-censoring inherent in survival data, we transform the censored survival times into \textit{pseudo-observations} \citep{Andersen:2010} for RMST estimation; see the detailed procedure in the Appendix (Section~A1). This transformation can be implemented using the R package \texttt{pseudo}. By treating pseudo-observations as outcomes, we effectively transform the time-to-event endpoint into a continuous response. Consequently, we can apply our propose methods for consistency assessment.

\section{Simulation}\label{sec:simu}
We conduct extensive simulations to evaluate the performance of our propose method for different types of endpoints in comparison with the one-step method of \citet{ko_criterion}.
 
\subsection{Setup}\label{subsec:setup}
Set the sample size of the region of interest to $n_r = 60$ and that of the complementary region to $n_{-r} = 340$, leading to $\rho_r=0.15$. This overall sample size and regional fraction represent a common case in practice. We consider four covariates, i.e., $p = 4$.  

To mimic a real trial setting, we assume that the support of ${\bm X}$ is bounded. Specifically, each component $X_s$ ($s = 1, 2, 3, 4$) is generated independently from a truncated normal distribution given by
\[
    f(x|a,b,\mu,\sigma) = \frac{\phi\left(\frac{x-\mu}{\sigma}\right)}{\sigma\left\{\Phi\left(\frac{b-\mu}{\sigma}\right)-\Phi\left(\frac{a-\mu}{\sigma}\right)\right\}}, \quad a \leq x\leq b,
\] 
where $\Phi$ and $\phi$ are the distribution function and density function of the standard normal variable, respectively. 
Throughout the simulations, the truncation bounds are fixed at $a = -3$ and $b = 3$, and the standard deviation is set to be $\sigma = 1.4$.
To introduce the distributional shift between the samples from the region $r$ of interest and its complementary regions, we set different values of the parameter $\mu$ of certain covariates associated with regions $r$ and $-r$.

The underlying model for the potential outcomes $Y_h(t)$ with $t\in \{-1, 1\}$ in region $h$ ($h=r, -r$) is specified as follows. 
For a continuous outcome in region $h$, assume
\[
Y_{h}(t) = 10 + 5\sum_{s=1}^{4}X_{s} + \kappa_{h}g(X_{1},X_{2})\frac{t+1}{2} + \varepsilon,
\]
where $\varepsilon \sim N(0,1)$ and $g$ is a known function depending only on $X_1$ and $X_2$. The CATE function of region $h$ is then $\Delta_{h}({\bm X}) = \kappa_{h}g(X_{1},X_{2})$, where $\kappa_{h}$ controls the scale of the CATE. For $g$, we consider three forms: (i) linear: $X_{1} + 0.5 X_{2}$, (ii) quadratic: $X_{1}^{2} - 1.61 + 0.5 (X_{2}^{2} - 1.61)$, and (iii) cubic: $X_{1}^{3} + 0.5 X_{2}^{3}$.
 
For a binary outcome, consider the logit-transformed probability of success in region $h$ given by 
\[
        \logit\, \mathbb{P}\{Y_{h}(t)=1\mid {\bm X}\} = -5 +  5(X_3+X_4) + \kappa_{h}(X_{1}+0.5X_{2}) \frac{t+1}{2}.  
\]
For a survival outcome, assume that the time to event in region $h$ follows the exponential distribution through 
\[
        Y_{h}(t) \sim \mathrm{Exp}(\lambda_{h, t})\ \textrm{with}\  \lambda_{h, t}  = X_3 + X_4 + \kappa_{h}(X_{1}+0.5X_{2}) \frac{t+1}{20}.  
\]
The censoring time is generated from a uniform distribution $\mathrm{U}[0,200]$, which yields a censoring rate of around 25\%. As discussed at the end of Section \ref{subsec:combined-consistency-assessment}, we will use \textit{pseudo-observations} \citep{Andersen:2010} to deal with censoring and work with RMST estimation based on  $Y^{*} = \min(Y, \tau)$, in which the restricted time $\tau$ is set to $100$.
Notice that for both binary and survival endpoint cases, the resulting CATE functions of region $h$ depend on all covariates, with the details of the analytical forms provided in the Appendix. The parameter $\kappa_{h}$ controls the scale of the CATE as in the continuous endpoint case.  

To introduce the discrepancy in the CATE between regions, we fix $\kappa_{-r} = 10$ and vary $\kappa_{r}$ across the set $\{0,2,4,6,8,10\}$, which leads to the ratio $\kappa_{r}/\kappa_{-r}$ spanning the discrete range $\{0,0.2,0.4,0.6,0.8,1\}$. The ratio of zero indicates a maximum difference in the CATE between regions and the ratio of one indicates identical CATEs between regions.

In our analysis, $X_{s}$ ($s=1,2,3,4$) are all discretized evenly into three levels based on their empirical 33\% and 66\% quantiles. The treatment $T$ is generated from a Bernoulli distribution with a success probability of 0.5.

Let $\mathrm{CP}^\mathrm{Ko}(q) = \prob(\widehat{\delta}_{r}/\widehat{\delta}_{-r} > q \vert \widehat{Z}> z_{\alpha})$ denote the consistency probability (CP) of region $r$ compared with its complementary region by the traditional one-step method \citep{ko_criterion}. Let $\mathrm{CP}^\mathrm{ts}(q^{(1)},q^{(2)}) = \prob( \text{Algorithm~\ref{algorithm:combined-consistency-assessment} claims consistency} \vert \widehat{Z}> z_{\alpha})$ denote the consistency probability of the same criterion obtained by our propose two-step method. Let $q=q^{(2)}=0.5$ and consider $q^{(1)}$ to be 0.5, 0.75 and 0.9, as proposed at the end of Section~\ref{subsec:consistency-covariate-shifts}. 
Throughout, the simulation is replicated $B=10,000$ times. 

\subsection{Results in the absence of covariate distributional shift}\label{sec:results-without-distribution-shift}
We first present simulation results in the case of no covariate distributional shift. That is, the covariate distributions are identical across the regions in the sense of  $F_r({X_s})=F_{-r}(X_s)$  (or a weaker version of $\E_{r}(X_{s}) = \E_{-r}(X_{s})$) for all $s=1,\ldots,4$. 

To introduce an ATE for the overall significance, we fix $\mu=0$ for $X_2$, $X_3$ and $X_4$ and tune the parameter $\mu$ for $X_1$ so that the powers of the MRCT (through the t-test) under the considered combinations of $\kappa_r$ and $\kappa_{-r}$ are all greater than $90\%$, as larger power makes it easier to demonstrate the improvement of CP by our propose method. 
We provide the resulting moments of $X_1$ used respectively in the linear, quadratic and cubic forms of $g$ and the associated regional ATEs in Table~\ref{tab:scenarios-without-distribution-shift} for the continuous endpoint and the resulting first moment of $X_1$ for the binary and survival endpoints in Table~\ref{tab:scenarios-binary-survival-without-distribution-shift} in the Appendix. Since the covariate distributions are identical, the regional ATE, $\delta_h$, is the same up to the scale parameter $\kappa_h$.

\begin{figure}[htbp]
    \centering
    
    \begin{subfigure}[b]{0.32\textwidth}
        \centering
        \includegraphics[width=\linewidth]{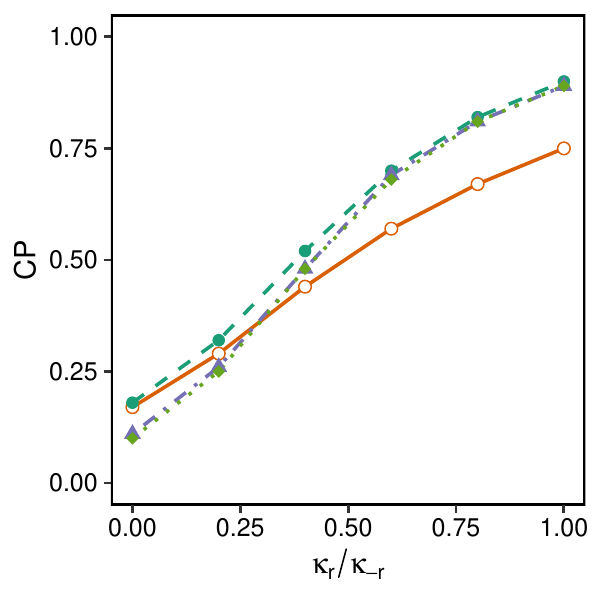}
        \caption{continuous -- linear} 
    \end{subfigure}
    \hfill
    \begin{subfigure}[b]{0.32\textwidth}
        \centering
        \includegraphics[width=\linewidth]{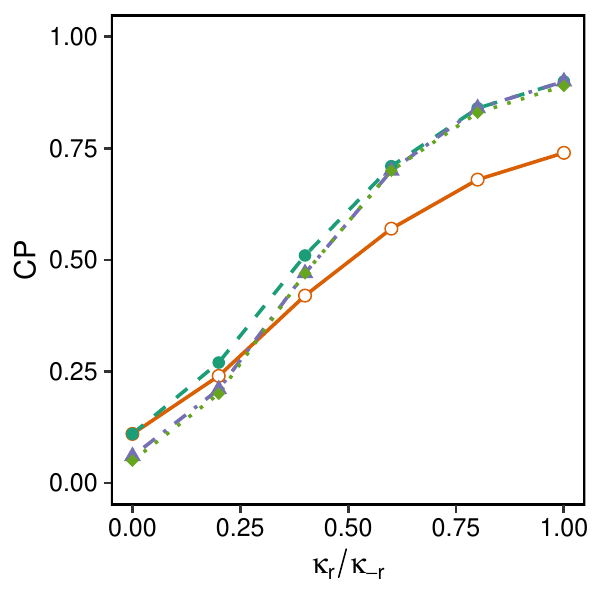}
        \caption{continuous -- quadratic} 
    \end{subfigure}
    \hfill
    \begin{subfigure}[b]{0.32\textwidth}
        \centering
        \includegraphics[width=\linewidth]{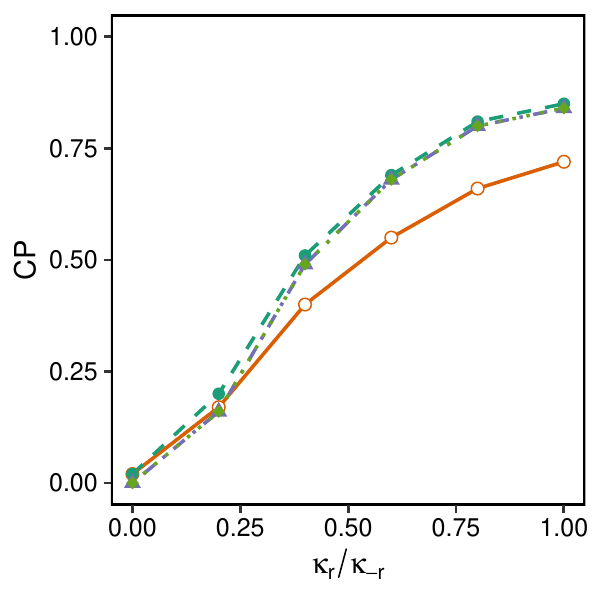}
        \caption{continuous -- cubic} 
    \end{subfigure}
    
    \vspace{1em} 
    
    \begin{subfigure}[b]{0.32\textwidth}
        \centering
        \includegraphics[width=\linewidth]{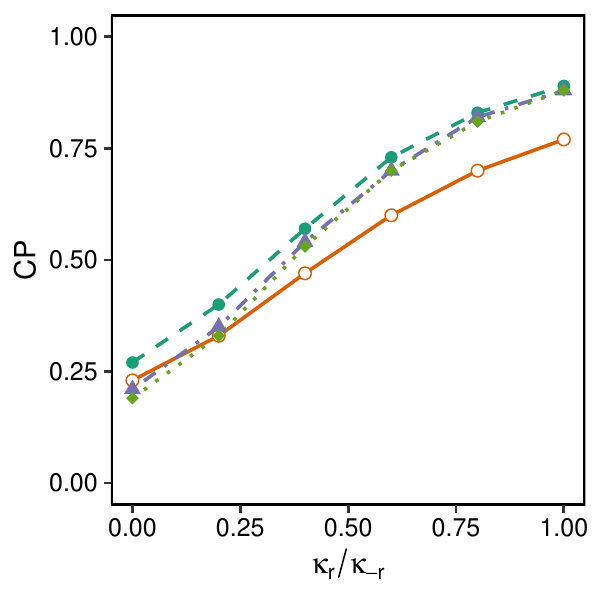}
        \caption{binary} 
    \end{subfigure}
    \hfill
    \begin{subfigure}[b]{0.32\textwidth}
        \centering
        \includegraphics[width=\linewidth]{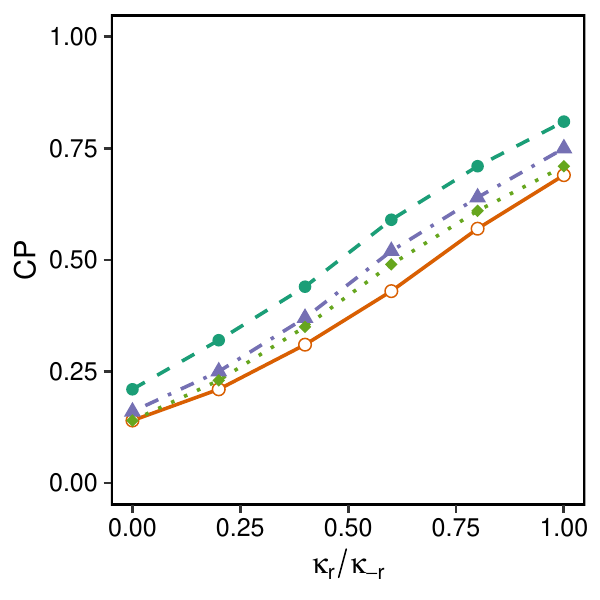}
        \caption{survival} 
    \end{subfigure}
    \hfill
    \begin{subfigure}[b]{0.32\textwidth}
        \centering
        \raisebox{2.5em}{\includegraphics[width=\linewidth]{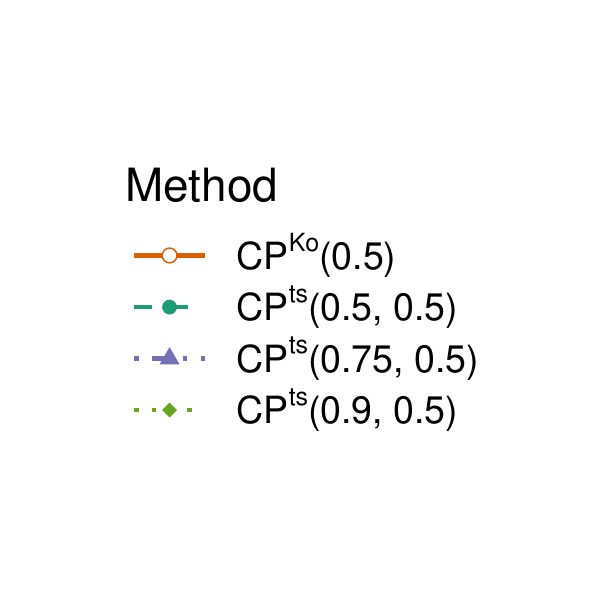}}
    \end{subfigure}
    
    \caption{Empirical CPs of region $r$ with respect to its complementary regions obtained by the one-step method and the proposed two-step method for various types of endpoints under different levels of discrepancy in regional CATEs (decreasing in $\kappa_r / \kappa_{-r}$) in the absence of covariate distributional shift}
    \label{fig:cp-without-distribution-shift}
\end{figure}

Under these settings, Figure~\ref{fig:cp-without-distribution-shift} shows the empirical CPs of region $r$ with respect to its complementary regions. As the ratio $\kappa_{r}/\kappa_{-r}$ increases, the discrepancy between the regional CATE functions diminishes gradually. Consequently, the CPs obtained by both methods increase accordingly. This trend aligns with our expectation, since, in the absence of a covariate distributional shift, the CATE function difference is the primary driver of inconsistency. 
Overall, the two-step method performs slightly more satisfactorily than the one-step method across all endpoints. The empirical CPs from the two-step method are generally lower when there is larger discrepancy between CATEs (when the ratio  $\kappa_{r}/\kappa_{-r}$ is small, e.g. less than 0.4); and are generally higher when there is smaller discrepancy between CATEs (when the ratio $\kappa_{r}/\kappa_{-r}$ is large, e.g. greater than 0.6). The results show that in the presence of covariate shift, our proposed two-step method is able to improve true positive rate (analogous to power) with well-controlled false positive rate (analogous to type I error) compared to the traditional one-step method by \citet{ko_criterion}, with the combination of ($q^{(1)},q^{(2)}$) = (0.9,0.5) demonstrating the most desirable operating characteristics as described earlier in the end of Section~\ref{subsec:consistency-covariate-shifts}.

\subsection{Results in presence of covariate distributional shift}\label{sec:results-with-distribution-shift}

Next, we consider scenarios with a covariate distributional shift.  
First, set $\mu = 0$ for $X_{3}$ and $X_{4}$ for both regions (so that they do not involve the covariate shift at all). Second, fix the parameters $\mu = 0$ for covariates $X_{s}$ ($s=1, 2$) in region $r$, leading to the first three moments $\E_{r}(X_{s})=0$, $\E_{r}(X_{s}^{2})=1.61$ and $\E_{r}(X_{s}^{3})=0$, as well as $\delta_r=0$. To introduce a non-zero ATE, we tune the parameters $\mu$ of $X_1$ and $X_2$ of the complementary region in the following two scenarios: (i) a shift in $X_1$ alone and (ii) shifts in both $X_1$ and $X_2$.
The values of the parameter $\mu$ are chosen so that the overall powers of the MRCT under the considered combinations of $\kappa_r$ and $\kappa_{-r}$ are all greater than $90\%$.
Recall that the CATE functions involve $X_1$ and $X_2$ for all endpoints.
We provide the resulting moments of $X_1$ and $X_2$, and the associated regional ATEs of the complementary region for the continuous endpoint in Table~\ref{tab:scenarios-with-distribution-shift} and the resulting first moments of $X_1$ and $X_2$ for the binary and survival endpoints in Table~\ref{tab:scenarios-binary-survival-with-distribution-shift} in the Appendix.

Figure~\ref{fig:cp-with-distribution-shift-i} shows the empirical CPs of region $r$ with respect to its complementary regions under scenario (i), where a covariate shift takes place in $X_1$ alone. We see that the two-step methods perform significantly better than the traditional one-step method in all cases. This enhanced performance is a crucial feature of our approach: even in the presence of a covariate shift, as long as the regional CATE functions are similar (when the ratio $\kappa_{r}/\kappa_{-r}$ is large, e.g. greater than 0.6), our proposed two-step methods produces much higher CP estimates than the one-step method. Conversely, when the regional CATE functions are dissimilar (when the ratio $\kappa_{r}/\kappa_{-r}$ is small, e.g. less than 0.4), the two-step methods tend to have similar performance to the one-step method, producing small CPs. Therefore, our proposed two-step method demonstrates desirable operating characteristics in the presence of distribution shift in treatment effect modifier in terms of improved true positive rate (analogous to power) when CATEs are similar between region $r$ and its complementary region ($-r$) and well-controlled false positive rate (analogous to type I error) when CATEs are dissimilar.
\begin{figure}[htbp]
    \centering
    
    \begin{subfigure}[b]{0.32\textwidth}
        \centering
        \includegraphics[width=\linewidth]{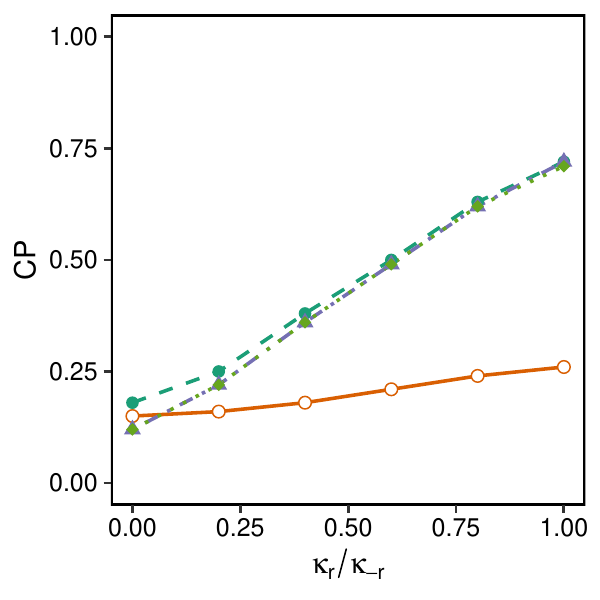}
        \caption{continuous -- linear} 
    \end{subfigure}
    \hfill
    \begin{subfigure}[b]{0.32\textwidth}
        \centering
        \includegraphics[width=\linewidth]{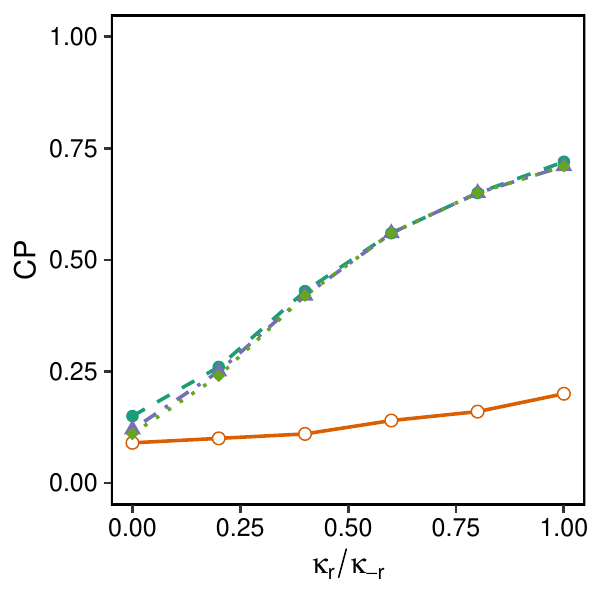}
        \caption{continuous -- quadratic} 
    \end{subfigure}
    \hfill
    \begin{subfigure}[b]{0.32\textwidth}
        \centering
        \includegraphics[width=\linewidth]{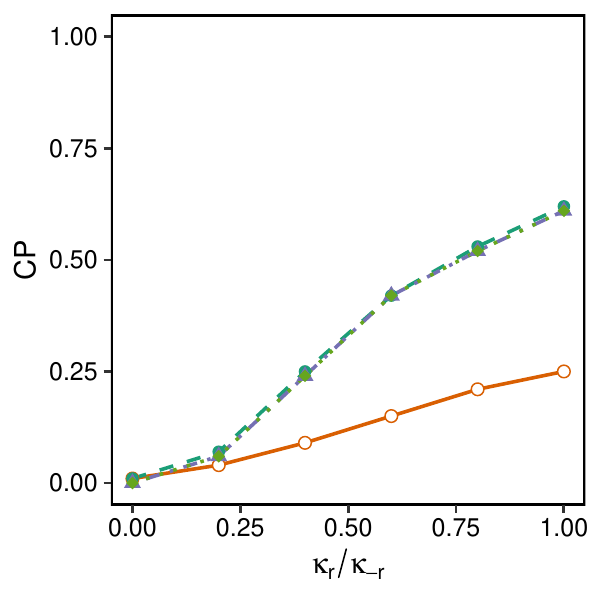}
        \caption{continuous -- cubic} 
    \end{subfigure}
    
    \vspace{1em} 
    
    \begin{subfigure}[b]{0.32\textwidth}
        \centering
        \includegraphics[width=\linewidth]{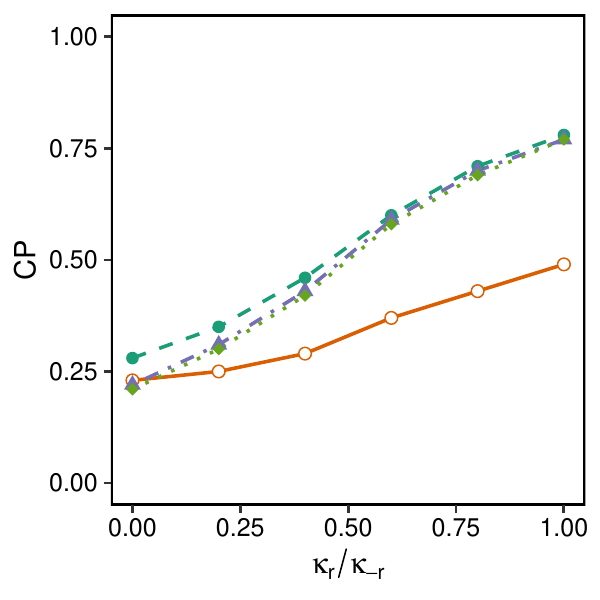}
        \caption{binary} 
    \end{subfigure}
    \hfill
    \begin{subfigure}[b]{0.32\textwidth}
        \centering
        \includegraphics[width=\linewidth]{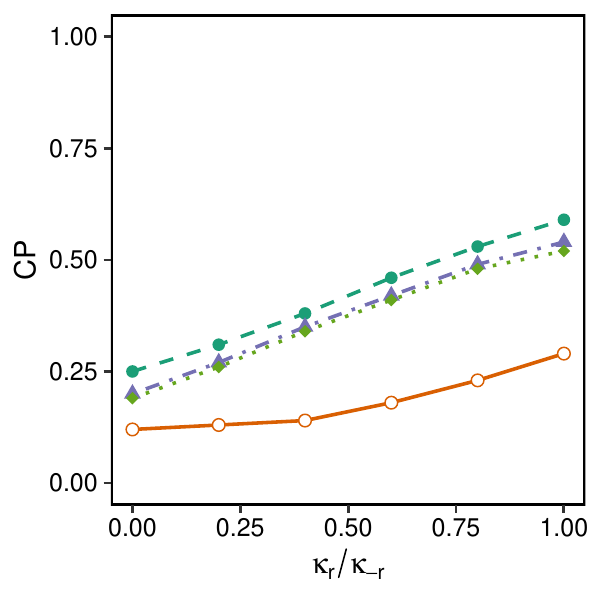}
        \caption{survival} 
    \end{subfigure}
    \hfill
    \begin{subfigure}[b]{0.32\textwidth}
        \centering
        \raisebox{2.5em}{\includegraphics[width=\linewidth]{figs/legend.pdf}}
    \end{subfigure}
    
    \caption{Empirical CPs of region $r$ with respect to its complementary regions obtained by the one-step method and the proposed two-step method under scenario (i), where a covariate shift takes place in $X_1$ alone, for various types of endpoints under different levels of discrepancy in regional CATEs (decreasing in $\kappa_r / \kappa_{-r}$)}
    \label{fig:cp-with-distribution-shift-i}
\end{figure}

\begin{figure}[htbp]
    \centering
    
    \begin{subfigure}[b]{0.32\textwidth}
        \centering
        \includegraphics[width=\linewidth]{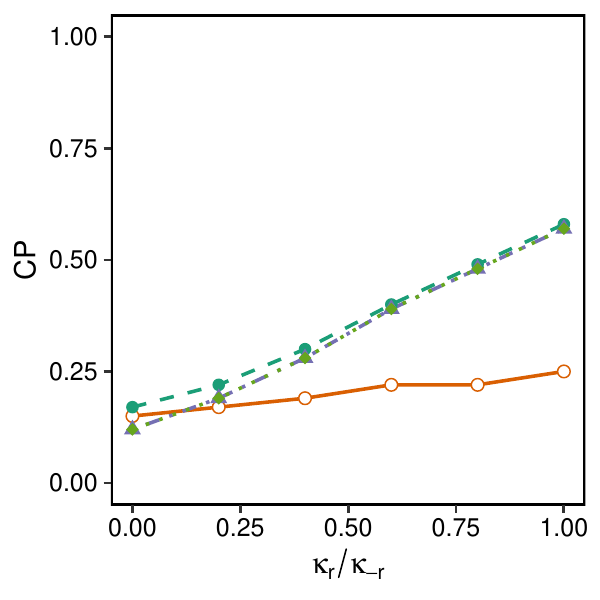}
        \caption{continuous -- linear} 
    \end{subfigure}
    \hfill
    \begin{subfigure}[b]{0.32\textwidth}
        \centering
        \includegraphics[width=\linewidth]{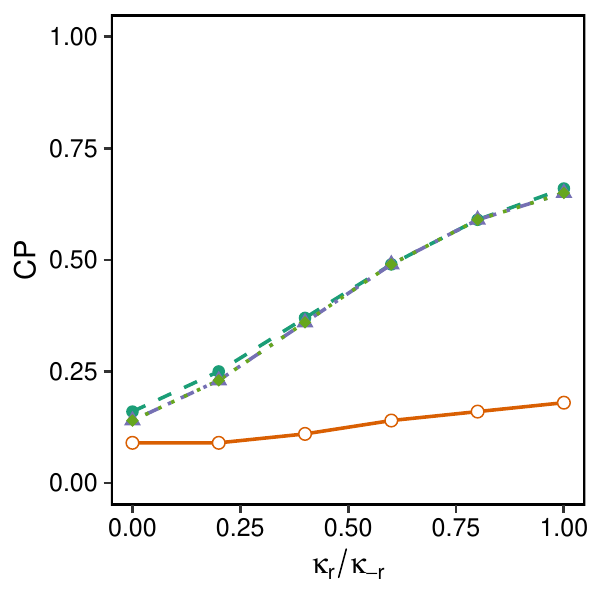}
        \caption{continuous -- quadratic} 
    \end{subfigure}
    \hfill
    \begin{subfigure}[b]{0.32\textwidth}
        \centering
        \includegraphics[width=\linewidth]{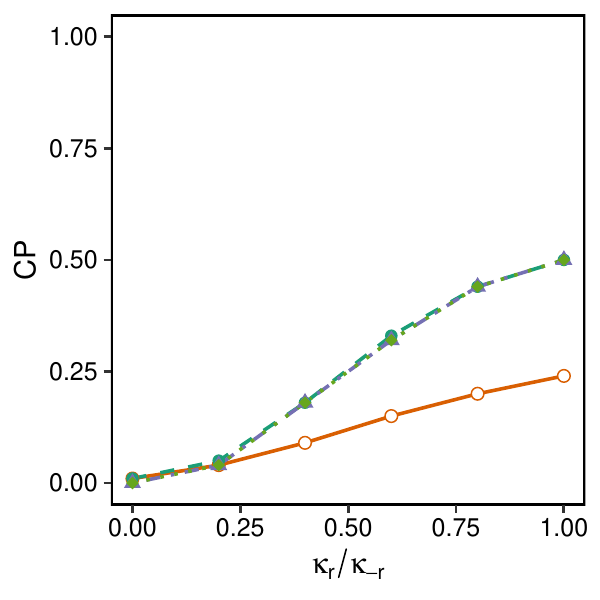}
        \caption{continuous -- cubic} 
    \end{subfigure}
    
    \vspace{1em} 
    
    \begin{subfigure}[b]{0.32\textwidth}
        \centering
        \includegraphics[width=\linewidth]{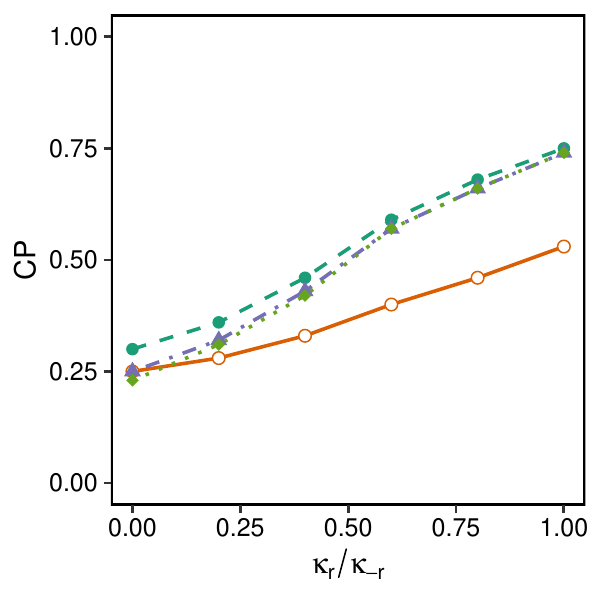}
        \caption{binary} 
    \end{subfigure}
    \hfill
    \begin{subfigure}[b]{0.32\textwidth}
        \centering
        \includegraphics[width=\linewidth]{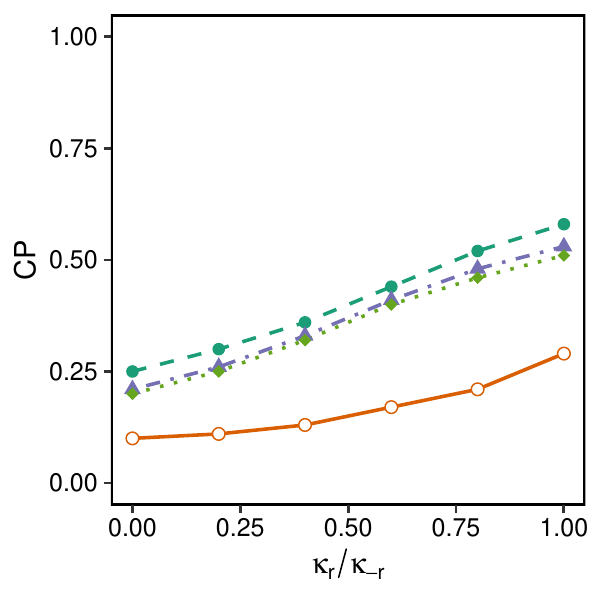}
        \caption{survival} 
    \end{subfigure}
    \hfill
    \begin{subfigure}[b]{0.32\textwidth}
        \centering
        \raisebox{2.5em}{\includegraphics[width=\linewidth]{figs/legend.pdf}}
    \end{subfigure}
    
    \caption{Empirical CPs of region $r$ with respect to its complementary regions obtained by the one-step method and the proposed two-step method under scenario (ii), where a covariate shift takes place in both $X_1$ and $X_2$, for various types of endpoints under different levels of discrepancy in regional CATEs (decreasing in $\kappa_r / \kappa_{-r}$)}
    \label{fig:cp-with-distribution-shift-continuous-ii}
\end{figure}

Figure~\ref{fig:cp-with-distribution-shift-continuous-ii} shows the empirical CPs of region $r$ with respect to its complementary regions under scenario (ii), where a covariate shift takes place in both $X_1$ and $X_2$. The findings are similar to those in scenario (i). That is, our proposed two-step method yields a similar low CP to that of the one-step method when CATEs are different across regions and rectifies consistency successfully with a much higher CP than the one-step method when CATEs are similar across regions.

Overall, we see $\mathrm{CP}^\mathrm{ts}(0.5,0.5)\ge\mathrm{CP}^\mathrm{Ko}(0.5)$, as expected. To control the false positive rate (analogous to type I error) when CATEs are different and to improve the true positive rate (analogous to power) when CATEs are similar, we recommend using $\mathrm{CP}^\mathrm{ts}(0.9,0.5)$ based on the evaluation of results from the simulations in Section~\ref{sec:results-without-distribution-shift} and \ref{sec:results-with-distribution-shift}. 
 
\section{Application to the BELIEVE trial}\label{sec:application}

We illustrate the application of the proposed two-step consistency assessment method to the BELIEVE trial \citep{Cappellini:2020}. This Phase III MRCT evaluated luspatercept in patients with transfusion-dependent $\beta$-thalassemia. The primary endpoint was a binary response (responder or non-responder) where responder was defined as a reduction in transfusion burden of at least $33\%$ from baseline, accompanied by a reduction of at least $2$ red-cell units, observed during weeks $13$ through $24$.

A total of 224 patients were assigned to the luspatercept group and 112 to the placebo group. The global trial results demonstrated a statistically significant treatment effect, with $21.4\%$ of patients in the luspatercept group achieving the primary endpoint compared to $4.5\%$ in the placebo group. This yielded an estimated odds ratio (OR) of $5.79$ ($95\%$ CI: $2.24, 14.97$). 

A subgroup analysis of $117$ Asian patients revealed a smaller, non-significant effect, with success rates of $14.8\%$ and $5.6\%$ for the luspatercept and placebo groups, respectively (OR: $3.06$; $95\%$ CI: $0.64, 14.63$). However, in a post hoc analysis based on a subgroup of the Asian population consisting of $68$ patients who had a baseline RBC transfusion burden (BTB) between $6$ and $15$ units per $24$ weeks prior to randomization (representing a lower range of BTB levels in the MRCT population), the responder rate ($21.7\%$ vs. $4.5\%$) and estimated OR ($5.83$) closely mirrored the global results. Clinical and regulatory reviews suggested that a higher BTB level corresponds to a smaller treatment benefit, considering BTB as a key effect modifier. This led to the conditional approval of luspatercept in China patients with transfusion-dependent $\beta$-thalassemia restricted to ``baseline transfusion burden $\leq 15$ units/24 weeks''.

Due to the unavailability of individual patient data from the trial, we simulate data that reproduce the summary results reported in the BELIEVE study. We assume that the CATE functions for Asian and non-Asian patients are identical. We use a logistic model to generate individual patient data and set the baseline transfusion burden (valued in the range $[6, 26]$ as reported in the BELIEVE trial) as the effect modifier.

The probability of achieving the primary endpoint $Y=1$ is modeled as:
\[ 
\logit\, \mathbb{P}(Y=1|\mathrm{BTB},T) = - 15 \cdot \frac{T+1}{2} + 1.3\cdot\mathrm{BTB}\cdot \frac{T+1}{2} + 2.7,
\] 
where $T$ is the treatment indicator. This specific logistic model parametrization is chosen because its resulting overall success rates, subgroup differences, and ORs align closely with those reported by the BELIEVE trial. A covariate shift in BTB is introduced between the Asian and Non-Asian populations. Specifically, we reconstruct data of the Asian subgroup with a large proportion (approximately $42\%$) of patients falling into the ``High BTB'' range ($15$--$26$ units). In contrast, the Non-Asian group is composed of a small proportion (approximately $18\%$) of patients falling into the ``High BTB'' range. In an effort to validate our data simulating process, we are able to reproduce the results reported by the BELIEVE trial with the above logistic model and our setup of the distributional shift in BTB. CPs by the traditional one-step method and our proposed two-step method are calculated using simulated data with $10,000$ replications. The results reveal a clear difference between the traditional one-step and our proposed two-step consistency assessment methods.

The overall power of the MRCT is 99\%. The one-step method, $\mathrm{CP}^\mathrm{Ko}(0.5)$, leads to an estimated consistency probability of 24\%. 
In contrast, the estimated consistency probabilities from the two-step methods, $\mathrm{CP}^\mathrm{ts}(0.5,0.5)$, $\mathrm{CP}^\mathrm{ts}(0.75,0.5)$, and $\mathrm{CP}^\mathrm{ts}(0.9,0.5)$, are 70\%, 69\% and 69\%, respectively. These results support the review opinion that the treatment effect could indeed be consistent across regions which was not revealed by the traditional one-step method due to shift in BTB among regions. More importantly, our proposed two-step method for consistency assessment is able to reveal the underlying consistent effect across regions in the presence of covariate shifts leading to the right claim.

\section{Discussion}\label{sec:disc}

The primary objective of an MRCT is to demonstrate the consistency of the treatment effect across all participating regions. In this paper, we proposed innovative methods for evaluating regional consistency that extend the widely adopted criterion initially introduced by \citet{MHLW:2007}. A key advantage of our approach is its ability to rectify false conclusions of inconsistency due to covariate shifts in important treatment effect modifiers. Our proposed method can be readily extended to non-inferiority (in which the null hypotheis becomes $H_0: \delta+M=0$ where $M$ is the non-inferiority margin) by applying a shift of size $M$ to $\widehat{\delta}$ in $\widehat{Z}$ and to $\widehat{\delta}_r$, $\widehat{\delta}_{-r}$, and $\widehat{\delta}^\ast_{r,s}$ in (\ref{eq:conditional-events}) and (\ref{eq:CATE-adjusted-event}), respectively.
%
Despite the promising numerical performance demonstrated in the numerical studies, a rigorous theoretical analysis is still warranted to provide guidance on issues such as the choice of $q^{(1)}$ and $q^{(2)}$ as thresholds for claiming consistency.
%



\section*{Acknowledgements}

The research was supported by a fund of the Merck \& Co., Inc. (MSD) R\&D (China) Co., LTD. The authors would like to thank Drs. William Wang, Shuping Jiang, Ping Yang from the MSD for their insightful discussions during the preparation of this manuscript. 

\bibliographystyle{apalike}
\bibliography{ConMRCT}

\section*{Appendix}
\renewcommand{\thesection}{A}
\renewcommand{\thefigure}{A\arabic{figure}}
\renewcommand{\thetable}{A\arabic{table}}

\subsection{Computation of pseudo-observations of RMST}
A primary challenge in survival analysis is the presence of right-censored data, where the event of interest is not observed for all subjects. When estimating the CATE functions based on the RMST, this censoring must be addressed. A common and effective method is to transform the censored time-to-event data into \textit{pseudo-observations} \citep{Andersen:2010}, which can then be used in standard regression frameworks.

The pseudo-observation approach generates a set of uncensored values that are (asymptotically) unbiased for the true RMST of each subject. The procedure is based on a jackknife (or ``leave-one-out'') statistic as follows.

\begin{enumerate}
	\item[(i)] \textbf{Estimate RMST based on the full data:} First, an estimate of the RMST up to time $\tau$, denoted $\widehat{\theta}$, is computed for the full dataset of $n$ subjects. This is done by using a nonparametric method that accounts for censoring, typically by integrating the Kaplan-Meier estimate of the survival function, $\widehat{S}(u)$. That is, 
	\[
	\widehat{\theta} = \int_0^{\tau} \widehat{S}(u) du.
	\]
	
	\item[(ii)] \textbf{Estimate RMST based on the leave-one-out data:} For each subject $i$ (from $1$ to $n$), the dataset is modified by removing subject $i$. A new RMST estimate, $\widehat{\theta}_{(i)}$, is then computed in the same way as in (i) from this reduced dataset of $n-1$ subjects.
	
	\item[(iii)] \textbf{Compute pseudo-observations:} The pseudo-observation $\widehat{\theta}_i^{\mathrm{po}}$ for subject $i$ is obtained as:
	\[
	\widehat{\theta}_i^{\mathrm{po}} = n \widehat{\theta} - (n-1)\widehat{\theta}_{(i)}.
	\]
\end{enumerate}

\subsection{CATE functions for binary and survival endpoints}\label{suppsec:CATE-function-binary-survival}

For binary endpoint, the CATE function of region $h$ ($h=r, \ -r$) is 
\[
\Delta_{h}(\bm{X}) = \frac{1}{1 + \exp\left[-\left\{-5 + 5(X_3+X_4) + \kappa_{h}(X_{1}+0.5X_{2})\right\}\right]} - \frac{1}{1 + \exp\left[-\left\{-5 + 5(X_3+X_4)\right\}\right]}.
\]

For survival endpoint of the exponential distribution with parameter $\lambda$, the RMST at truncation point $\tau$ is $(1 - e^{-\lambda \tau})/\lambda$.
Therefore, given the parameter $\lambda_{h,t}$ defined in Section~4.1, 
the CATE of region $h$ is  
\[ \Delta_{h}(\bm{X}) 
= \frac{1 - \exp\left[-\tau \left\{ X_3 + X_4 + 0.1 \kappa_{h}(X_{1}+0.5X_{2})  \right\}\right]}{X_3 + X_4 + 0.1\kappa_{h}(X_{1}+0.5X_{2}) } - \frac{1 - \exp\left\{-\tau \left(X_3 + X_4  \right)\right\}}{X_3 + X_4}.
\]

\subsection{Details of the setup of scenarios for Sections~\ref{sec:results-without-distribution-shift} and \ref{sec:results-with-distribution-shift}}
\begin{table}[htbp]
	\centering
	\caption{The moments of $X_1$ for the continuous endpoint under the linear, quadratic and cubic forms of $g$ (whose underlying parameters $\mu$ are 0.8, 1.4 and 0.5, respectively) such that the powers of the MRCT (through the t-test) under the considered combinations of $\kappa_r$ and $\kappa_{-r}$ are all greater than $90\%$. The associated regional ATE $\delta_{h}$ ($h=r, -r$), where $\mu=0$ for $X_2$, $X_3$ and $X_4$, is given in the last column.}
	\label{tab:scenarios-without-distribution-shift}
	\begin{tabular}{ccc}  %
		\hline
		$g$ & $X_{1}$ & $\delta_{h}$ \\
		\hline
		linear & $\E(X_{1})=0.64$ & $0.64\kappa_{h}$ \\ 
		quadratic & $\E(X_{1}^{2})=2.45$ & $0.84\kappa_{h}$ \\ 
		cubic & $\E(X_{1}^{3})=1.62$ & $1.62\kappa_{h}$ \\ 
		\hline
	\end{tabular}
\end{table}

\begin{table}[t]
	\centering
	\caption{The first moment of $X_1$ for the binary and survival endpoints (along with $\mu=0$ for $X_2$, $X_3$ and $X_4$) such that the powers of the MRCT (through the t-test) under the considered combinations of $\kappa_r$ and $\kappa_{-r}$ are all greater than $90\%$ (The associated regional ATEs are obtained numerically by integration of the CATEs given in Section~\ref{suppsec:CATE-function-binary-survival}.)}
	\label{tab:scenarios-binary-survival-without-distribution-shift}
	\begin{tabular}{ll}  %
		\hline
		endpoint & $\qquad\quad X_{1}$  \\
		\hline
		binary &  $ \E(X_{1})=0.408$ \\  
		survival & $\E(X_{1})=-0.791$ \\   
		\hline
	\end{tabular}
\end{table}

\begin{table}[htbp]
	\centering
	\caption{The moments of $X_1$ and $X_2$ of region $-r$ under the linear, quadratic and cubic forms of $g$ for the case of the continuous endpoint and two scenarios of covariate shifts (i.e., (i) a shift in $X_1$ alone and (ii) shifts in both $X_1$ and $X_2$) such that the powers of the MRCT (through the t-test) under the considered combinations of $\kappa_r$ and $\kappa_{-r}$ are all greater than $90\%$. The resulting regional ATE $\delta_{-r}$ is given in the last column.}
	\label{tab:scenarios-with-distribution-shift}
	\begin{tabular}{lllc}  %
		\hline
		scenario & $\qquad\quad X_{1}$ & $\qquad X_{2}$ & $\delta_{-r}$\\
		\hline
		linear (i) & $\E_{-r}(X_{1})=0.72$ & $\E_{-r}(X_{2})=0$ & $0.72\kappa_{-r}$ \\ 
		linear (ii) & $\E_{-r}(X_{1})=0.49$ & $\E_{-r}(X_{2})=0.49$ & $0.74\kappa_{-r}$ \\   
		quadratic (i) & $\E_{-r}(X_{1}^{2}-1.61)=0.94$ & $\E_{-r}(X_{2}^{2}-1.61)=0$ & $0.94\kappa_{-r}$ \\   
		quadratic (ii) & $\E_{-r}(X_{1}^{2}-1.61)=0.64$ & $\E_{-r}(X_{2}^{2}-1.61)=0.64$ & $0.96\kappa_{-r}$ \\  
		cubic (i) & $\E_{-r}(X_{1}^{3})=1.96$ & $\E_{-r}(X_{2}^{3})=0$ & $1.96\kappa_{-r}$ \\  
		cubic (ii) & $\E_{-r}(X_{1}^{3})=1.29$ & $\E_{-r}(X_{2}^{3})=1.29$ & $1.94\kappa_{-r}$ \\  
		\hline
	\end{tabular}
\end{table}

\begin{table}[htbp]
	\centering
	\caption{The first moments of $X_1$ and $X_2$ of region $-r$ for the cases of binary and survival endpoints under two scenarios of covariate shifts ((i) a shift in $X_1$ alone and (ii) shifts in both $X_1$ and $X_2$) such that the powers of the MRCT (through the t-test) under the considered combinations of $\kappa_r$ and $\kappa_{-r}$ are all greater than $90\%$. (The associated regional ATEs are obtained numerically by integration of the CATEs given in Section~\ref{suppsec:CATE-function-binary-survival}.) }
	\label{tab:scenarios-binary-survival-with-distribution-shift}
	\begin{tabular}{lll}  %
		\hline
		scenario & $\qquad X_{1}$ & $\qquad X_{2}$  \\
		\hline
		binary (i) & $\E_{-r}(X_{1})=0.408$ & $\E_{-r}(X_{2})=0$  \\   
		binary (ii) & $\E_{-r}(X_{1})=0.245$ & $\E_{-r}(X_{2})=0.164$  \\  
		survival (i) & $\E_{-r}(X_{1})=-0.866$ & $\E_{-r}(X_{2})=0$  \\ 
		survival (ii) & $\E_{-r}(X_{1})=-0.486$ & $\E_{-r}(X_{2})=-0.407$  \\    
		\hline
	\end{tabular}
\end{table}

\begin{table}[t] 
	\centering
	\caption{Empirical CPs of region $r$ with respect to its complementary regions obtained by the one-step method and the proposed two-step method for various types of endpoints under different levels of discrepancy in regional CATEs (decreasing in $\kappa_r / \kappa_{-r}$) in the absence of covariate distributional shift}
	\label{tab:cp-without-distribution-shift}
	
	\begin{tabular}{ll *{6}{c}}
		\hline 
		\multirow{2}{*}{scenario} & \multirow{2}{*}{method} & \multicolumn{6}{c}{  $\kappa_r / \kappa_{-r}$} \\
		\cmidrule(lr){3-8} 
		& & 0.0 & 0.2 & 0.4 & 0.6 & 0.8 & 1.0 \\
		\hline 
		
		continuous -- linear & $\mathrm{CP}^\mathrm{Ko}(0.5)$ & .17 & .29 & .44 & .57 & .67 & .75 \\
		& $\mathrm{CP}^\mathrm{ts}(0.5,0.5)$ & .18 & .32 & .52 & .70 & .82 & .90 \\
		& $\mathrm{CP}^\mathrm{ts}(0.75,0.5)$ & .11 & .26 & .48 & .69 & .81 & .89 \\
		& $\mathrm{CP}^\mathrm{ts}(0.9,0.5)$ & .10 & .25 & .48 & .68 & .81 & .89 \\
		\hline 
		
		continuous -- quadratic & $\mathrm{CP}^\mathrm{Ko}(0.5)$ & .11 & .24 & .42 & .57 & .68 & .74 \\
		& $\mathrm{CP}^\mathrm{ts}(0.5,0.5)$ & .11 & .27 & .51 & .71 & .84 & .90 \\
		& $\mathrm{CP}^\mathrm{ts}(0.75,0.5)$ & .06 & .21 & .47 & .70 & .84 & .90 \\
		& $\mathrm{CP}^\mathrm{ts}(0.9,0.5)$ & .05 & .20 & .47 & .70 & .83 & .89 \\
		\hline 
		
		continuous -- cubic & $\mathrm{CP}^\mathrm{Ko}(0.5)$ & .02 & .17 & .40 & .55 & .66 & .72 \\
		& $\mathrm{CP}^\mathrm{ts}(0.5,0.5)$ & .02 & .20 & .51 & .69 & .81 & .85 \\
		& $\mathrm{CP}^\mathrm{ts}(0.75,0.5)$ & .00 & .16 & .49 & .68 & .80 & .84 \\
		& $\mathrm{CP}^\mathrm{ts}(0.9,0.5)$ & .00 & .16 & .49 & .68 & .80 & .84 \\
		\hline 
		
		binary & $\mathrm{CP}^\mathrm{Ko}(0.5)$ & .23 & .33 & .47 & .60 & .70 & .77 \\
		& $\mathrm{CP}^\mathrm{ts}(0.5,0.5)$ & .27 & .40 & .57 & .73 & .83 & .89 \\
		& $\mathrm{CP}^\mathrm{ts}(0.75,0.5)$ & .21 & .35 & .54 & .70 & .82 & .88 \\
		& $\mathrm{CP}^\mathrm{ts}(0.9,0.5)$ & .19 & .33 & .53 & .70 & .81 & .88 \\
		\hline 
		
		survival & $\mathrm{CP}^\mathrm{Ko}(0.5)$ & .14 & .21 & .31 & .43 & .57 & .69 \\
		& $\mathrm{CP}^\mathrm{ts}(0.5,0.5)$ & .21 & .32 & .44 & .59 & .71 & .81 \\
		& $\mathrm{CP}^\mathrm{ts}(0.75,0.5)$ & .16 & .25 & .37 & .52 & .64 & .75 \\
		& $\mathrm{CP}^\mathrm{ts}(0.9,0.5)$ & .14 & .23 & .35 & .49 & .61 & .71 \\
		\hline 
	\end{tabular}
\end{table}

\begin{table}[t] 
	\centering
	\caption{Empirical CPs of region $r$ with respect to its complementary regions obtained by the one-step method and the proposed two-step method under scenario (i), where a covariate shift takes place in $X_1$ alone, for various types of endpoints under different levels of discrepancy in regional CATEs (decreasing in $\kappa_r / \kappa_{-r}$)}
	\label{tab:cp-with-distribution-shift-i}
	
	\begin{tabular}{ll *{6}{c}}
		\hline
		\multirow{2}{*}{scenario} & \multirow{2}{*}{method} & \multicolumn{6}{c}{ $\kappa_r / \kappa_{-r}$} \\
		\cmidrule(lr){3-8} 
		& & 0.0 & 0.2 & 0.4 & 0.6 & 0.8 & 1.0 \\
		\hline        
		continuous -- linear & $\mathrm{CP}^\mathrm{Ko}(0.5)$ & .15 & .16 & .18 & .21 & .24 & .26 \\
		& $\mathrm{CP}^\mathrm{ts}(0.5,0.5)$ & .18 & .25 & .38 & .50 & .63 & .72 \\
		& $\mathrm{CP}^\mathrm{ts}(0.75,0.5)$ & .12 & .22 & .36 & .49 & .62 & .72 \\
		& $\mathrm{CP}^\mathrm{ts}(0.9,0.5)$ & .12 & .22 & .36 & .49 & .62 & .71 \\
		\hline        
		continuous -- quadratic & $\mathrm{CP}^\mathrm{Ko}(0.5)$ & .09 & .10 & .11 & .14 & .16 & .20 \\
		& $\mathrm{CP}^\mathrm{ts}(0.5,0.5)$ & .15 & .26 & .43 & .56 & .65 & .72 \\
		& $\mathrm{CP}^\mathrm{ts}(0.75,0.5)$ & .12 & .25 & .42 & .56 & .65 & .71 \\
		& $\mathrm{CP}^\mathrm{ts}(0.9,0.5)$ & .11 & .24 & .42 & .56 & .65 & .71 \\
		\hline
		continuous -- cubic & $\mathrm{CP}^\mathrm{Ko}(0.5)$ & .01 & .04 & .09 & .15 & .21 & .25 \\
		& $\mathrm{CP}^\mathrm{ts}(0.5,0.5)$ & .01 & .07 & .25 & .42 & .53 & .62 \\
		& $\mathrm{CP}^\mathrm{ts}(0.75,0.5)$ & .00 & .06 & .24 & .42 & .52 & .61 \\
		& $\mathrm{CP}^\mathrm{ts}(0.9,0.5)$ & .00 & .06 & .24 & .42 & .52 & .61 \\
		\hline        
		binary  & $\mathrm{CP}^\mathrm{Ko}(0.5)$ & .23 & .25 & .29 & .37 & .43 & .49 \\
		& $\mathrm{CP}^\mathrm{ts}(0.5,0.5)$ & .28 & .35 & .46 & .60 & .71 & .78 \\
		& $\mathrm{CP}^\mathrm{ts}(0.75,0.5)$ & .22 & .31 & .43 & .59 & .70 & .77 \\
		& $\mathrm{CP}^\mathrm{ts}(0.9,0.5)$ & .21 & .30 & .42 & .58 & .69 & .77 \\
		\hline        
		survival & $\mathrm{CP}^\mathrm{Ko}(0.5)$ & .12 & .13 & .14 & .18 & .23 & .29 \\
		& $\mathrm{CP}^\mathrm{ts}(0.5,0.5)$ & .25 & .31 & .38 & .46 & .53 & .59 \\
		& $\mathrm{CP}^\mathrm{ts}(0.75,0.5)$ & .20 & .27 & .35 & .42 & .49 & .54 \\
		& $\mathrm{CP}^\mathrm{ts}(0.9,0.5)$ & .19 & .26 & .34 & .41 & .48 & .52 \\
		\hline
	\end{tabular}
\end{table}

\begin{table}[t] 
	\centering
	\caption{Empirical CPs of region $r$ with respect to its complementary regions obtained by the one-step method and the proposed two-step method under scenario (ii), where a covariate shift takes place in both $X_1$ and $X_2$, for various types of endpoints under different levels of discrepancy in regional CATEs (decreasing in $\kappa_r / \kappa_{-r}$)}
	\label{tab:cp-with-distribution-shift-continuous-ii}
	\begin{tabular}{ll *{6}{c}}
		\hline         \multirow{2}{*}{scenario} & \multirow{2}{*}{method} & \multicolumn{6}{c}{ $\kappa_r / \kappa_{-r}$} \\
		\cmidrule(lr){3-8} 
		& & 0.0 & 0.2 & 0.4 & 0.6 & 0.8 & 1.0 \\
		\hline        
		continuous -- linear & $\mathrm{CP}^\mathrm{Ko}(0.5)$ & .15 & .17 & .19 & .22 & .22 & .25 \\
		& $\mathrm{CP}^\mathrm{ts}(0.5,0.5)$ & .17 & .22 & .30 & .40 & .49 & .58 \\
		& $\mathrm{CP}^\mathrm{ts}(0.75,0.5)$ & .12 & .19 & .28 & .39 & .48 & .57 \\
		& $\mathrm{CP}^\mathrm{ts}(0.9,0.5)$ & .12 & .19 & .28 & .39 & .48 & .57 \\
		\hline        
		continuous -- quadratic & $\mathrm{CP}^\mathrm{Ko}(0.5)$ & .09 & .09 & .11 & .14 & .16 & .18 \\
		& $\mathrm{CP}^\mathrm{ts}(0.5,0.5)$ & .16 & .25 & .37 & .49 & .59 & .66 \\
		& $\mathrm{CP}^\mathrm{ts}(0.75,0.5)$ & .14 & .23 & .36 & .49 & .59 & .65 \\
		& $\mathrm{CP}^\mathrm{ts}(0.9,0.5)$ & .14 & .23 & .36 & .49 & .59 & .65 \\
		\hline
		continuous -- cubic & $\mathrm{CP}^\mathrm{Ko}(0.5)$ & .01 & .04 & .09 & .15 & .20 & .24 \\
		& $\mathrm{CP}^\mathrm{ts}(0.5,0.5)$ & .01 & .05 & .18 & .33 & .44 & .50 \\
		& $\mathrm{CP}^\mathrm{ts}(0.75,0.5)$ & .00 & .04 & .18 & .32 & .44 & .50 \\
		& $\mathrm{CP}^\mathrm{ts}(0.9,0.5)$ & .00 & .04 & .18 & .32 & .44 & .50 \\
		\hline        
		binary  & $\mathrm{CP}^\mathrm{Ko}(0.5)$ & .25 & .28 & .33 & .40 & .46 & .53 \\
		& $\mathrm{CP}^\mathrm{ts}(0.5,0.5)$ & .30 & .36 & .46 & .59 & .68 & .75 \\
		& $\mathrm{CP}^\mathrm{ts}(0.75,0.5)$ & .25 & .32 & .43 & .57 & .66 & .74 \\
		& $\mathrm{CP}^\mathrm{ts}(0.9,0.5)$ & .23 & .31 & .42 & .57 & .66 & .74 \\
		\hline        
		survival  & $\mathrm{CP}^\mathrm{Ko}(0.5)$ & .10 & .11 & .13 & .17 & .21 & .29 \\
		& $\mathrm{CP}^\mathrm{ts}(0.5,0.5)$ & .25 & .30 & .36 & .44 & .52 & .58 \\
		& $\mathrm{CP}^\mathrm{ts}(0.75,0.5)$ & .21 & .26 & .33 & .41 & .48 & .53 \\
		& $\mathrm{CP}^\mathrm{ts}(0.9,0.5)$ & .20 & .25 & .32 & .40 & .46 & .51 \\
		\hline
	\end{tabular}
\end{table}

\end{document}